\documentclass[aps,prl,twocolumn,superscriptaddress]{revtex4-2}

\usepackage{graphicx}
\usepackage{dcolumn}
\usepackage{bm}
\usepackage{amsmath}
\usepackage{amssymb}
\usepackage{latexsym}
\usepackage{epsfig}
\usepackage{amsbsy}
\usepackage{array}
\usepackage{amssymb}
\usepackage{setspace}
\usepackage{bm}
\usepackage{epstopdf}
\usepackage{subfigure}
\usepackage{color}

\def\sint{\ifmmode{- \!\!\!\!\!\! \int}
    \else{\hbox{$- \!\!\!\! \int \ $}}\fi}

\begin{document}

\title{A general interpretation of nonlinear connected time crystals: quantum self-sustaining combined with quantum synchronization}

\author{Song-hai Li}
\affiliation{College of Sciences, Northeastern University, Shenyang 110819, China}
\affiliation{Research Center for Quantum Physics, Northeastern University, Shenyang 110819, China}
\author{Najmeh Es’haqi-Sani}
\affiliation{Instituto de Ciencia de Materiales de Madrid, Consejo Superior de Investigaciones Científicas, Calle Sor Juana Inés de la Cruz, 3, 28049 Madrid, Spain}
\affiliation{Instituto de Nanociencia y Materiales de Aragon, CSIC-Universidad de Zaragoza, 50009 Zaragoza, Spain}
\author{Xingli Li}
\affiliation{Department of Physics, The Chinese University of Hong Kong, Shatin, New Territories, Hong Kong, China}
\affiliation{Lanzhou Center for Theoretical Physics, Key Laboratory of Theoretical Physics of Gansu Province, Key Laboratory of Quantum Theory and Applications of MoE, Gansu Provincial Research Center for Basic Disciplines of Quantum Physics, Lanzhou University, Lanzhou 730000, China}
\author{Wenlin Li}
\email{liwenlin@mail.neu.edu.cn}
\email{wenlin.li@unicam.it}
\affiliation{College of Sciences, Northeastern University, Shenyang 110819, China}
\affiliation{Research Center for Quantum Physics, Northeastern University, Shenyang 110819, China}

\date{\today}
\begin{abstract}
Although classical nonlinear dynamics suggests that sufficiently strong nonlinearity can sustain oscillations, quantization of such model typically yields a time-independent steady state that respects time-translation symmetry and thus precludes time-crystal behavior. We identify dephasing as the primary mechanism enforcing this symmetry, which can be suppressed by intercomponent phase correlations. Consequently, a sufficient condition for realizing a continuous time crystal is a nonlinear quantum self-sustaining system exhibiting quantum synchronization among its constituents. As a concrete example, we demonstrate spontaneous oscillations in a synchronized array of van der Pol oscillators, corroborated by both semiclassical dynamics and the quantum Liouville spectrum. These results reduce the identification of time crystals in many-body systems to the evaluation of only two-body correlations and provide a framework for classifying uncorrelated time crystals as trivial.
\end{abstract}
\maketitle
The concept of time crystals, initially proposed by Wilczek as temporal analogues of spatial crystals, posits the spontaneous breaking of time-translation symmetry in systems at thermodynamic equilibrium~\cite{Wilczek2012}. Early proposals, however, were constrained by no-go theorems, which precluded persistent rotation in ground states or thermal equilibrium for a broad class of systems~\cite{Watanabe2015}. The turning point appears in periodically driven systems, known as Floquet time crystals~\cite{Sacha2015}, where the observables oscillate at multiples of the driving period, thereby breaking \textit{discrete} spontaneous time-translation invariance. This paradigm has since emerged as a mainstream paradigm, supported by extensive theoretical and experimental validation~\cite{Zaletel2023}.

Interest in continuous time crystals (CTCs) was reignited by Iemini~\textit{et al.}~\cite{Iemini2018} through the construction of a boundary time crystal characterized by the surface of a bulk system that self-organizes into a time-periodic pattern, with a period that depends only on its coupling constants. Crucially, while the theorem that the density matrix of a system in contact with a single thermal reservoir attains a time-independent steady state is still recognized in mathematics, the relaxation time required for the system to reach this steady state will diverge in the thermodynamic limit where the number of particles in the ensemble tends to infinity. 

Current theoretical frameworks for CTCs primarily rely on dynamical simulation or spectral analysis (e.g., identifying pure imaginary Liouville spectra)~\cite{Iemini2018,Li2024,Hajdu2024}, leaving their microscopic origins unresolved. A preliminary synthesis of extant work suggests that CTCs originate from nonlinearity, drawing parallels to quantum self-sustaining systems, a prototypical class of nonlinear quantum systems~\cite{Lee2013,Roulet2018}. While both CTCs and quantum self-sustaining systems exhibit limit cycle attractors and self-sustained oscillations in the corresponding classical limits, a critical distinction emerges quantum mechanically, time-translation symmetry is preserved in quantum self-sustaining systems, unlike that in CTCs.

\begin{figure}[]
\centering
\includegraphics[width=3in]{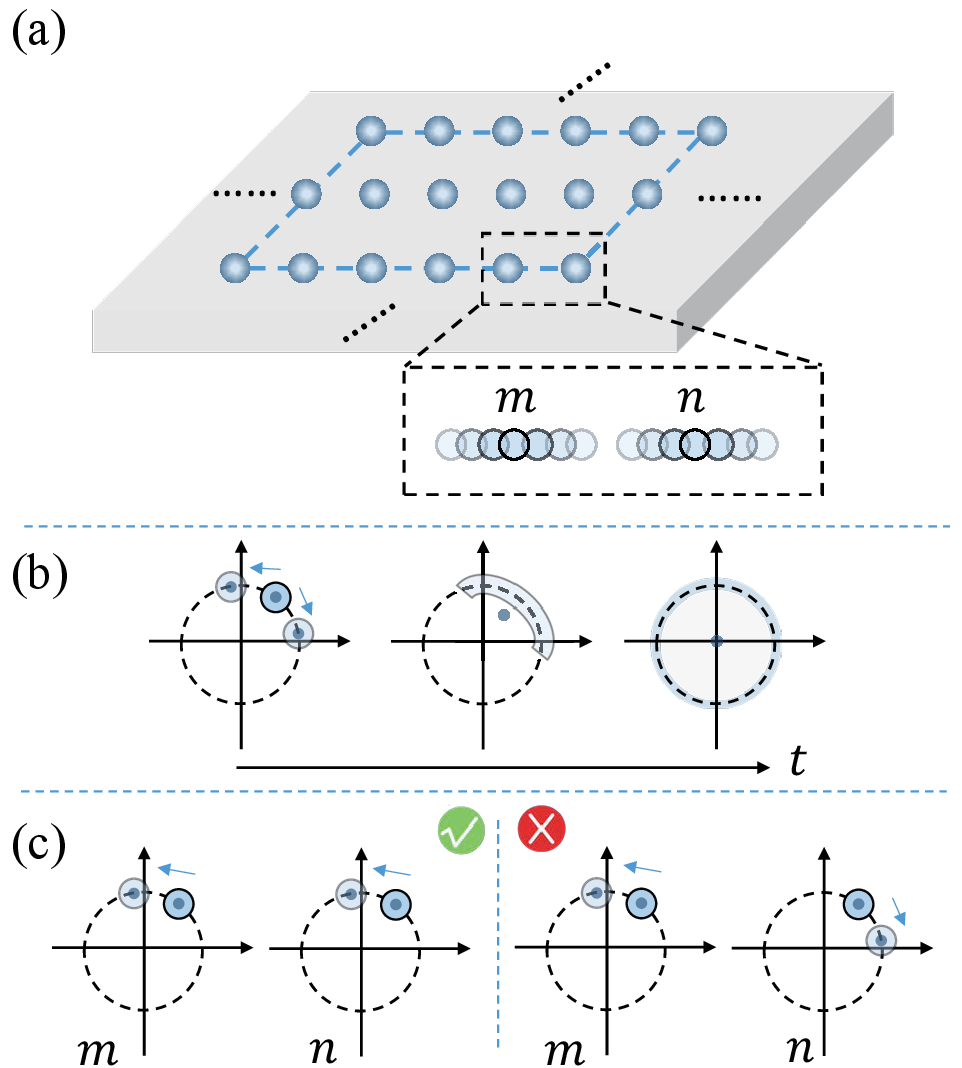}
\caption{(a): Schematic of the studied time crystal. The crystal surface is modeled as a 2D array of van der Pol oscillators, while the bulk serves as a common reservoir. Long-range interactions among surface oscillators emerge through their coupling to this shared reservoir. (b): Mechanism by which a classical self-sustained oscillator restores time-translation symmetry upon quantization. Every point on the classical limit cycle is a steady state, so phase noise diffuses freely, spreading the probability distribution along the trajectory. The blue region depicts the probability distribution, with the blue dot marking the mean value of the field operator ($\langle \hat{a}\rangle$). (c): Synchronization-induced breaking of time-translation symmetry. Left: Phase diffusion in the same direction preserves synchronization, yielding a valid steady state and thus time-translation symmetry. Right: Diffusion in opposite directions disrupts synchronization; the synchronization mechanism suppresses this process, breaking time-translation symmetry.
\label{fig:1}}
\end{figure}

In this Letter, we identify quantum synchronization among subsystems as the pivotal mechanism inducing the phase transition of a quantum self-sustaining system from trivial steady-state phase into a CTC phase. We further propose a general framework for constructing CTCs that leverages nonlinear interactions and long-range interactions. These two requirements have been extensively explored in their respective fields before converging in the study of CTCs. The mechanism is prevalent across diverse systems, including arrays of optomechanics~\cite{Ludwig2013,Aspelmeyer2014}, magnomechanics~\cite{Zu2024}, and van der Pol (VdP) oscillators~\cite{Lee2013,Lee2014}. To demonstrate, we analyze multiple CTC configurations composed of VdP oscillators as shown in Fig.~\ref{fig:1}(a).

Our analysis begins by positing quantum self-sustaining systems as a prerequisite for time crystals. To transcend formal mathematical arguments, we elucidate the physical mechanism that reverts such systems to trivial steady states by extending the analysis of Navarrete-Benlloch~\cite{Navarrete-Benlloch2017}. Consider a system exhibiting a limit cycle in the classical limit. Here, time is parameterized by phase, and time-translation symmetry breaking maps onto the dispersion of phase coherence: the limit cycle consists of a continuum of steady states differing only in phase, which we term degenerate states under dissipative dynamics. Quantum fluctuations induce transitions among these degenerate states.  As these states are equivalent under dissipation, no restorative mechanism exists, yielding undamped dynamics along the phase direction. Consequently, quantum fluctuations diffuse freely along the limit cycle (see Fig.~\ref{fig:1}(b)), resulting in a uniform phase distribution that averages out the phase evolution and eliminates time dependence. Thus, we identify the suppression of phase noise as the central mechanism for CTCs. Notably, our framework explains existing CTCs in the high-excitation limit and shows that systems lacking internal correlations cannot sustain time-crystal behavior. When quantum-level phase synchronization occurs between systems, a key distinction from the uncorrelated case arises: the phase-correlation constraint renders not all points on the limit cycle degenerate. For instance, as illustrated in Fig.~\ref{fig:1}(c), consider phase diffusion induced by quantum fluctuations on each limit cycle. With probability $1/2$, this diffusion proceeds in the same direction on both cycles, preserving the phase correlation between oscillators and thus being permitted. Dynamically, the steady states before and after such diffusion remain degenerate, with no corrective mechanism. Conversely, in the remaining $1/2$ of cases, opposite-directional diffusion introduces a phase mismatch, which the quantum synchronization mechanism counters, restoring phase alignment. Dynamically, these diffused states are excluded as steady states, with dissipation driving their elimination. Overall, this halves the effective free diffusion rate, thereby doubling the timescale for reaching the steady state. This can be straightforwardly generalized. For $N$ subsystems all achieving quantum synchronization, the probability of phase diffusion unaffected by correlations reduces to $1/2^{N-1}$, approaching zero in the thermodynamic limit. Notably, the uniform phase-diffusion state remains the system's steady state, but the time required to reach complete phase diffusion diverges, analogous to previously constructed boundary time crystals~\cite{Iemini2018}.

We interpret all nonlinear time crystals as arising from the interplay of two mechanisms: nonlinearity excites subsystems into quantum self-sustaining states, while quantum synchronization suppresses phase diffusion induced by quantum fluctuations, thereby prolonging the timescale to the steady state of complete phase diffusion, which diverges in the thermodynamic limit. This suggests that CTCs are not rare, as self-sustaining systems and quantum synchronization have been observed in diverse physical platforms, exemplified by the van der Pol (vdP) oscillators. As a detailed explanation, we now consider a quantum ensemble consisting of vdP oscillator, described by the Hamiltonian $H=\sum_n (\omega_n \hat{a}_n^\dagger\hat{a}_n+\Omega_n \hat{a}+\Omega_n^*\hat{a}_n^\dagger)$  ($\hbar=1$). Here $\hat{a}_n$ represents the annihilation operator for the $n$-th oscillator with resonance frequency $\omega_n$, and $\Omega_n$ is the complex driving amplitude. The dynamics are governed by the Lindblad master equation~\cite{Lee2014}
\begin{equation}
\begin{split}
\dot{\rho}=&-i[H,\rho]+\sum_n  \left(\kappa_1\mathcal{L}[a^\dagger_n]\rho+\kappa_2 \mathcal{L}[a^2_n]\rho\right)\\
&+\dfrac{\mu_{mn}}{\mathcal{N}}\sum_{m<n}\mathcal{L}[a_m-a_n]\rho,
\end{split}
\label{eq:master QLE}
\end{equation}
where $\mathcal{L}[\hat{o}]\rho=2\hat{o}\rho\hat{o}^\dagger-(\hat{o}^\dagger\hat{o}\rho+\rho\hat{o}^\dagger\hat{o})$ denotes the Liouvillian in Lindblad form for operator $\hat{o}$. The first Lindblad term provides linear gain at rate $\kappa_1$, while the second term induces nonlinear dissipation at rate $\kappa_2$. For a single oscillator, their balance yields a time-independent quantum limit cycle, manifesting as a ring distribution in Wigner phase space~\cite{Lee2014,Li2021}. It admits an exact solution and has been studied in detail previously. The energy scale of the steady state is governed by the ratio $\kappa_1/\kappa_2$. For $\kappa_1 \gg \kappa_2$, the oscillator exhibits large-amplitude oscillations, corresponding to the classical limit, where Eq.~(\ref{eq:op classical QLE}) can be approximately mapped onto a stochastic Langevin equation. In contrast, for $\kappa_1 \ll \kappa_2$, the steady state excites only a few quanta above the ground state, rendering the classical treatment invalid~\cite{Lee2013}. The last term represents long-range dissipative coupling with coupling strength $\mu_{mn}$, achievable through reservoir engineering, with $\mathcal{N}$ a normalization factor ensuring finite interaction strength in the thermodynamic limit $N \rightarrow \infty$, where $N$ is the number of oscillators. We first examine a one-dimensional chain with periodic boundary conditions, wherein the coupling decays exponentially with distance: $\mu_{mn}=\mu\lambda_{nm}(\gamma) =\mu (1-\delta_{mn})e^{-\gamma (d_{mn} - 1)}$ and $\mathcal{N}(\gamma) = \sum_{m=1}^{N} \lambda_{1m}+1$. Here, $\gamma$ is the attenuation coefficient, and $d_{mn}$ denotes the dimensionless lattice distance between sites $m$ and $n$. 	We apparently obtain $\lambda_{mn}(0) = 1$ and $\mathcal{N}(0) = N$, corresponding to an all-to-all coupled ensemble. Here we consider the case that the oscillators are identical ($\forall_n\,\omega_n:=\omega_m $), then any nonzero coupling strength induces quantum synchronization among oscillators.
\begin{figure}[]
\centering
\includegraphics[width=3in]{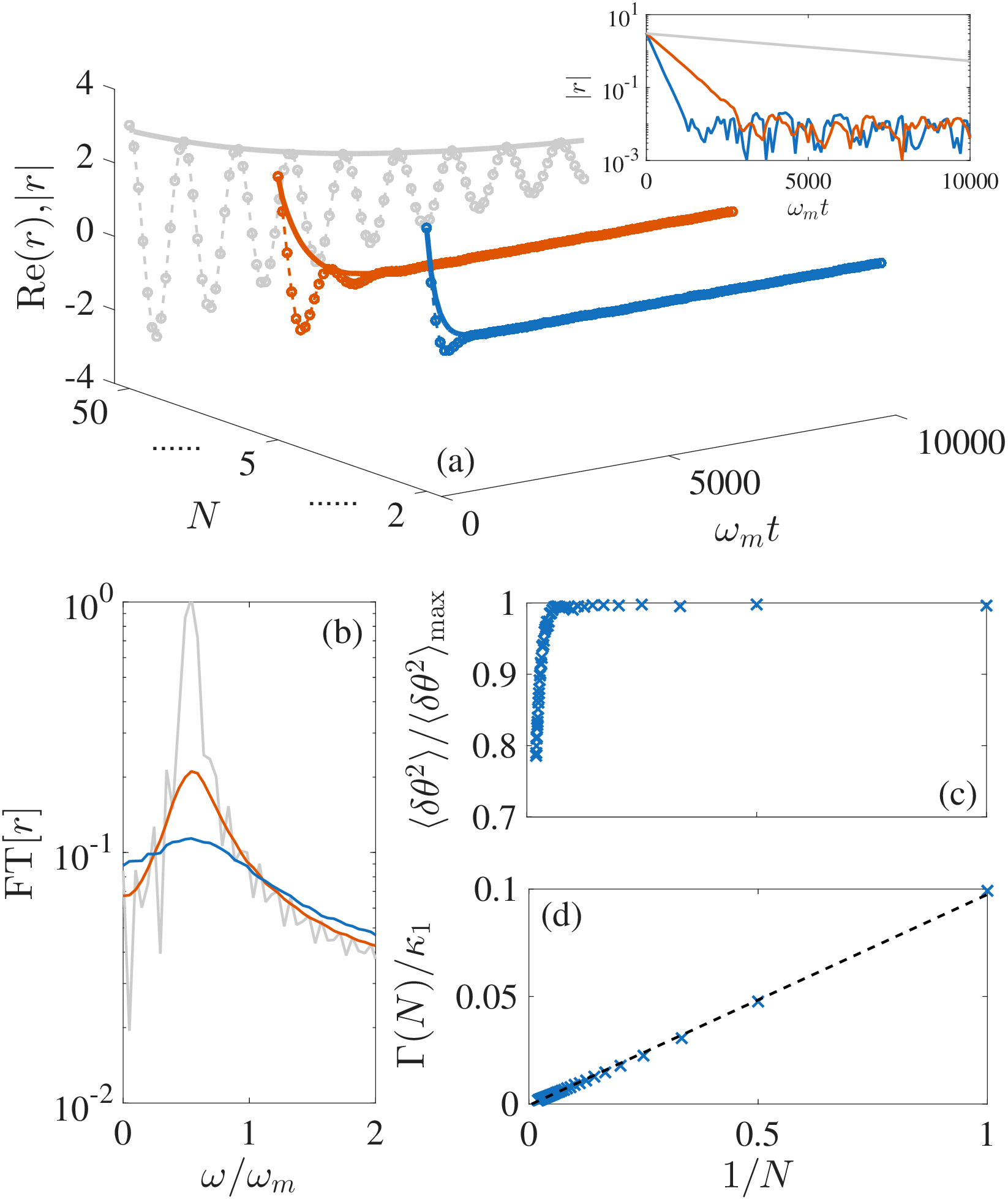}
\caption{(a) Time evolution of the order parameter $  r  $ and its amplitude for various $  N  $. The inset shows the amplitude on a logarithmic scale.(b) Fourier transform of the order parameter, with $  N  $ values color-coded as in (a). (c) and (d) Phase fluctuations and effective amplitude dissipation rate versus $  1/N  $. Here,  we set $\omega_m = 1$ as the unit, and the other parameters are $  \kappa_1=0.1$, $\kappa_2 = 0.005$, and $\mu=0.3$ and $\gamma=0$. Panels (a) and (b) derive from $100,000$ stochastic simulations using the Langevin equation~\eqref{eq:op classical QLE}; panels (c) and (d) from $50,000$.
\label{fig:2}}
\end{figure}

We first consider the semiclassical regime $\kappa_1 \gg \kappa_2$ mapping the system's quantum dynamics to stochastic trajectories in classical phase space governed by the Langevin equation~\cite{Lee2014,Li2021}: 
\begin{equation}
\begin{split}
\dot{a}_n=&-i\omega_n a_n+a_n(\kappa_1-2\kappa_2\vert a_n\vert^2)+\sqrt{3\kappa_1+2\kappa_2}a^{in}_n\\&+\sum_{m=1}^{N}\left[ \dfrac{\mu_{mn}}{\mathcal{N}(\gamma)}(a_m-a_n)+\sqrt{\dfrac{\mu_{mn}}{\mathcal{N}(\gamma)}}c^{in}_{mn}\right],
\end{split}
\label{eq:op classical QLE}
\end{equation}
where $a^{in}_n$ and $c^{in}_{mn}=-c^{in}_{nm}$ are the Gaussian vacuum input noises acting on the $n$-th oscillator, with the correlation function $\langle a^{in*}_n(t)a^{in}_n(t')\rangle=\langle a^{in}_n(t')a^{in*}_n(t)\rangle=\delta(t-t')$ and $\langle [c^{in}_{mn}(t)-c^{in}_{nm}(t)]^* [c^{in}_{mn}(t')-c^{in}_{nm}(t')]\rangle=\delta(t-t')$~\cite{Li2021,book1}. This approach circumvents the need for a large Hilbert space, enabling simulations for large $N$. More importantly, it provides an intuitive phase-space picture of the phase diffusion mechanism mentioned above.
\begin{figure}[]
\centering
\begin{subfigure}{}
    \includegraphics[width=1.7in]{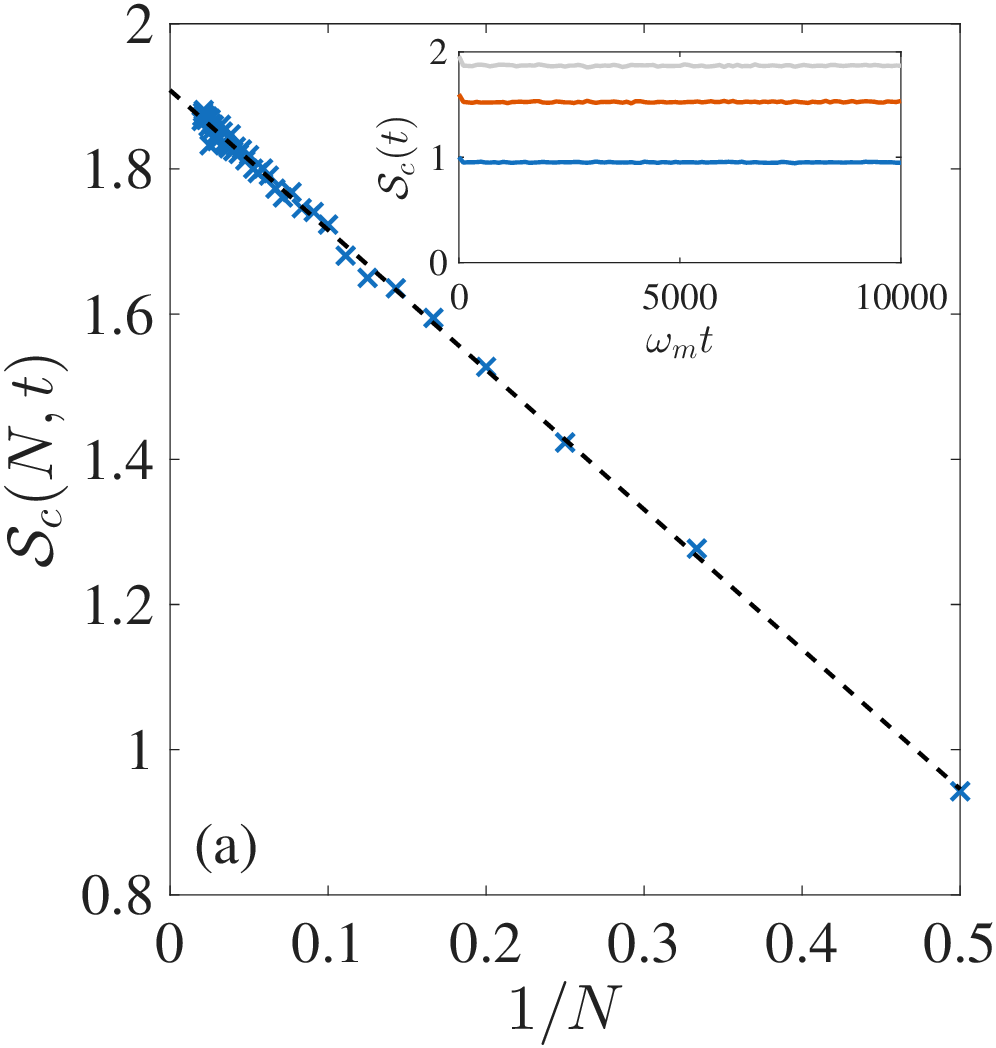}
\end{subfigure}
\begin{subfigure}{}
    \includegraphics[width=1.52in]{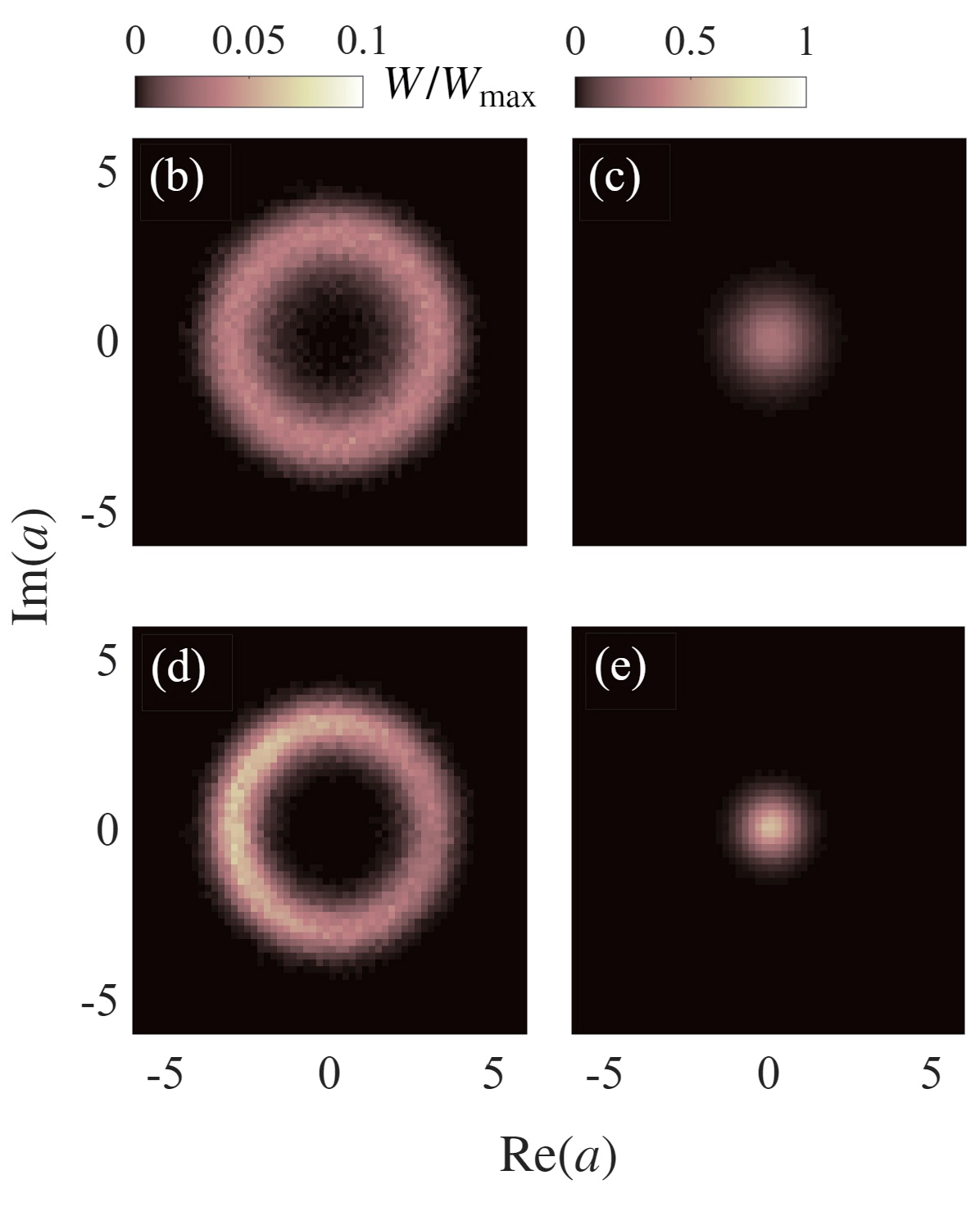}
\end{subfigure}
\caption{(a): Multi-body synchronization measure $\mathcal{S}_c(N,t)$ as a function of $1/N$. The inset shows its time evolution for $N=2$, $5$, and $50$, with color coding as in Fig.~\ref{fig:2}. (b) and (d): Phase-space probability distributions for oscillator 1 for $N=2$ and $5$. (c) and (e) Corresponding probability distributions of error-mode which is defined in Eq.~\eqref{eq:Sc N}.  Here we set $\omega_m t=10000$ and  all parameters match those in Fig.~\ref{fig:2}.}
\label{fig:3}
\end{figure}

The order parameter characterizing the breaking of time-translation symmetry in the system is defined as $ r = \frac{1}{N} \sum_i \langle a_i \rangle $, which temporal evolution is employed phenomenologically investigate the time-crystal properties. As shown in Fig.~\ref{fig:2}(a), $ \vert r\vert $ decays gradually until $ \vert r\vert = 0 $ and $ \dot{\vert r\vert } = 0 $, reflecting the restoration of time-translation symmetry. The corresponding spectral analysis in Fig.~\ref{fig:2}(b) reveals that the peak sharpens with increasing particle number. Sub-figure~(c) illustrates the link between amplitude decay and dephasing by calculating phase fluctuations $\langle \delta^2\theta\rangle$, where $\theta=\arg[a_1]$: for small particle numbers, the oscillator phases degenerate nearly to a uniform distribution, causing the limit cycle to vanish; as the particle number grows, phase fluctuations are markedly suppressed, enabling sustained oscillations over extended times. The upper-right inset of Fig.~\ref{fig:2}(a) shows that this amplitude decay is well fitted by a power law $\exp(-\Gamma t)$ with a effective dissipation rate $ \Gamma $. Although dissipation persists, the rate $ \Gamma $ decreases with increasing $ N $. Fig.~\ref{fig:2}(d) displays dimensionless $\Gamma/\kappa_1$ with fits yielding $0.1/N+\mathcal{O}(10^{-5})$. In the thermodynamic limit $ N \to \infty $, $ \Gamma \to 0 $, implying divergent dissipation timescales and hence spontaneous breaking of time-translation symmetry.

\begin{figure}[]
\centering
\includegraphics[width=3in]{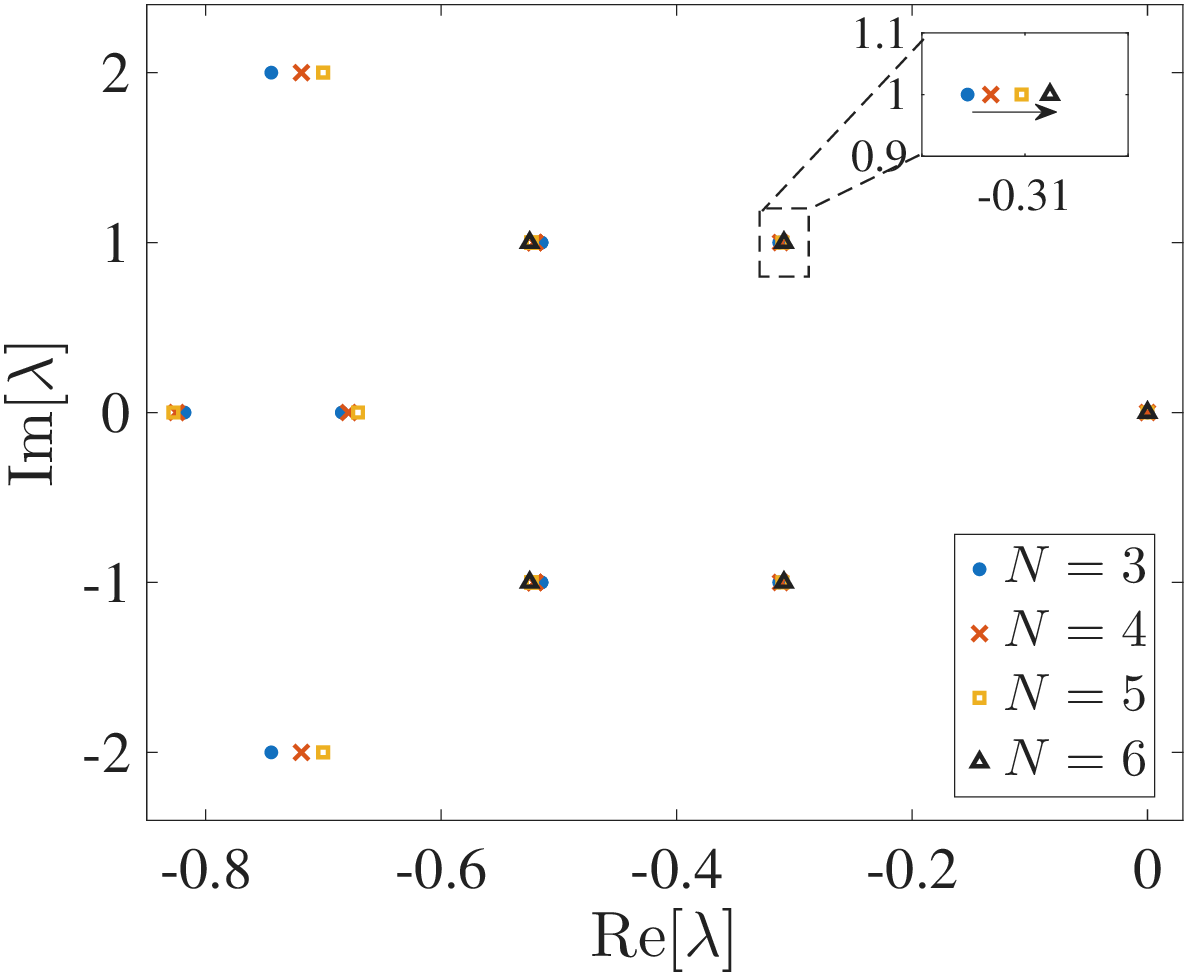}
\caption{The eigenvalues $\lambda$ of the Liouvillian.  The insets show an enlargement over the eigenvalues with the largest real part. The parameters here is  $\kappa_2 = 0.2$, and the other parameters match those in Fig.~\ref{fig:2}.
\label{fig:4}}
\end{figure}
We further elucidate how temporal many-body effects emerge from quantum phase synchronization, i.e., the collective alignment of oscillator phases at the microscopic level.  In our model, the nonlinearity of the vdP oscillator does not induce mode competition, yielding equal energy across subsystems~\cite{Li2021,Kemiktarak2014}. Accordingly, we characterize phase correlations using the complete synchronization criterion in terms of the conjugate quadratures $\hat{q}$ and $\hat{p}$ ($[\hat{q},\hat{p}]=i$), circumventing ambiguities inherent in phase-operator definitions.  Higher-order synchronization, incorporating cross-correlations between two sub-systems $1$ and $2$, is quantified via the Mari's measure: $\mathcal{S}_c(t)=\langle \hat{q}_-(t)^2+\hat{p}_-(t)^2\rangle^{-1}$, wherein nonlocal effects arise from the error operator $\hat{o}_-=(\hat{o}_1-\hat{o}_2)/\sqrt{2}$ for $\hat{o}\in\{q,p\}$~\cite{Mari2013}. For the many-body case, we extend Mari's criterion as follows: we select an arbitrary oscillator (e.g., oscillator $1$, without loss of generality due to the symmetry of all-to-all coupling) as system 1, and the mean field formed by the remaining oscillators (i.e., the collective mode $a$) as system $2$. Under this definition, Mari's measure is generalized as: $\mathcal{S}_c(N,t)=\langle \hat{q}_{N-}(t)^2+\hat{p}_{N-}(t)^2\rangle^{-1}$, with
\begin{equation}
\begin{split}
\hat{o}_{{N-}_{o=q,p}}= \dfrac{1}{\sqrt{2}}\left[\hat{o}_1(t)-\frac{\sum_{i=2}^{N}\hat{o}_i(t)}{N-1}\right]^2.
\end{split}
\label{eq:Sc N}
\end{equation}
Notably, in the classical limit where the two effective systems surpass the Heisenberg uncertainty principle bounds and attain exact equality ($\hat{a}_1=\frac{1}{N-1}\sum_{i \neq 1}\hat{a}_i$), Eq.~\eqref{eq:master QLE} reduces to a mean-field master equation. In that case, oscillator 1 couples not to the other oscillators but to the average field it generates~\cite{Jin2016}. Solutions to this equation describe self-sustained oscillations in the classical regime. Note that for uncorrelated subsystems in identical states, corresponding to classical synchronization without quantum correlations, the measure satisfies $\mathcal{S}_c(N)\propto\frac{1}{1+1/N} $. In contrast, Fig.~\ref{fig:3}(a) plots the synchronization measure $\mathcal{S}_c $ versus the number of subsystems $N$ at $\omega_m t=10000$, where correlations induce a linear scaling with $1/N$. This indicates that correlations strongly suppress relative errors among the oscillators and drive the emergence of quantum synchronization. The y-intercept reflects the achievable synchronization limit, approaching the quantum upper bound of $\mathcal{S}_c(N\rightarrow\infty)\rightarrow 2$, as oscillator 1 remains subject to the Heisenberg uncertainty principle in our definition. The inset of Fig.~\ref{fig:3}(a) demonstrates that $\mathcal{S}_c$ reaches high values for $N=2$, $5$, and $50$, even with substantial fluctuations in the oscillators ($\Delta  q_i\simeq \Delta  p_i\simeq 10$). Moreover, correlations form rapidly in our system, enabling $\mathcal{S}_c$ to stabilize at a constant value. Figures~\ref{fig:3}(b) to~(e) display the corresponding phase-space distributions. Panels~(b) and~(d) show those for subsystem 1 at $  N=2  $ and $50$, respectively. Although the Hamiltonian drives rotation at frequency $\omega_m$, the nearly symmetric distribution in (b) renders this rotation unobservable, whereas the asymmetry in~(d) manifests in the evolution of observable quantities. Panels~(c) and~(e) depict the error distributions, which follow a Gaussian form with a narrower spread for $N=50$.



The preceding analysis applies to limit-cycle radii with $  |\langle a_i \rangle| > 3  $. For smaller cycles, however, single-quantum creation or annihilation strongly perturbs the system, rendering the quantum model distinct from a classical noisy oscillator. Fortunately, few energy excitations suffice to capture full quantum property in a truncated Hilbert space. In particular, spontaneous time-translation symmetry breaking is characterized by the Liouvillian spectrum of master equation~\eqref{eq:master QLE}. In this framework, the real part of the Liouvillian eigenvalues corresponds to the effective oscillation frequency of the mode, while the imaginary part denotes its effective dissipation rate. A time crystal thus manifests through non-zero purely imaginary eigenvalues in the Liouvillian spectrum. Results are shown in Fig.~4, computed up to $  N=6  $ due to numerical limitations. All spectra exhibit a zero eigenvalue, corresponding to the asymptotic steady state. The remaining eigenvalues have negative imaginary parts, indicating gradual dissipation of their modes. However, increasing $  N  $ shifts these eigenvalues rightward, progressively reducing the effective dissipation rate. Although exact spectral decomposition for larger $N$ is infeasible,  optimistic qualitative extrapolation suggests that, in the fully quantum regime at $N\rightarrow \infty$, the eigenvalues approach the imaginary axis, analogous to the semiclassical case, yielding an ideal time crystal. 

Finally, our mechanism enables qualitative resolution of several key questions concerning time crystals, obviating quantitative computations. First, regarding rigidity: Synchronization inherently establishes a potential well for errors (e.g., a washboard potential)~\cite{Lee2013,Li2021,Cheng2023}, conferring robustness to moderate noise and subsystem heterogeneities. Thus, time crystals formed under this mechanism are intrinsically rigid. Second, are long-range interactions essential? As quantum synchronization can arise in systems with only nearest-neighbor couplings~\cite{Mari2013}, provided sufficient strength, our constructed time crystals do not require long-range interactions. Third, can arbitrary interactions induce time crystals? No; for example, two-body anti-synchronization precludes multibody time-crystal formation~\cite{Lee2013,Cheng2023}. This implies that the viability of time crystals in many identical systems can be assessed solely via two-body dynamics.

In summary, we propose a general mechanism for constructing continuous-variable time crystals, comprising two key elements: (i) each subsystem forms a nonlinearly induced quantum self-sustained oscillator, and (ii) quantum synchronization emerges among subsystems. The underlying principle is that quantum self-sustained systems restore time-translation symmetry via phase diffusion, whereas inter-subsystem correlations in synchronized ensembles suppress phase diffusion, reinstating time-dependent oscillations in the thermodynamic limit. We illustrate this mechanism using coupled vdP oscillators, with applicability extending to platforms like optomechanics and magnomechanics. Analyses span semiclassical and fully quantum regimes. Semiclassically, simulations for large subsystem counts yield an inverse scaling of the amplitude's effective dissipation rate with $N$, extrapolating to $\Gamma(N)\rightarrow0$ at $N\rightarrow\infty$. In the quantum regime, limited to small $N$, the Liouvillian spectrum reveals a decreasing magnitude of real parts with increasing $N$. This framework obviates exhaustive numerical searches for time crystals and clarifies contentious issues, including mean-field-based time crystals~\cite{Yang2025} and divergent interpretations in  Ref.~\cite{Li2024} and its comment~\cite{Navarrete-Benlloc2024}. Moreover, it elucidates the role of nonlocal correlations in quantum synchronization, justifying their inclusion in synchronization measures~\cite{Mari2013,Li2017}.

\begin{acknowledgements}
W. L. is supported by the National Natural Science Foundation of China (Grant No. 12304389) and Scientific Research Foundation of NEU (Grant No. 01270021920501*115). X. Li is supported by the National Natural Science Foundation of China (Grant No. 12247101), the Fundamental Research Funds for the Central Universities (Grant No. lzujbky-2025-jdzx07), the Natural Science Foundation of Gansu Province (No.25JRRA799), and the ‘111 Center’ under Grant No. B20063.
\end{acknowledgements}

\end{document}